\documentclass[a4paper]{jpconf}
\usepackage{graphicx}
\begin{document}
\title{$\bar K^*$ meson in nuclear matter}

\author{Laura Tolos$^1$, Raquel Molina$^2$, E. Oset$^2$ and A. Ramos$^3$}

\address{$^1$Theory Group. KVI. University of Groningen,
Zernikelaan 25, 9747 AA Groningen, The Netherlands}

\address{$^2$Instituto de F{\'\i}sica Corpuscular (centro mixto CSIC-UV),
Institutos de Investigaci\'on de Paterna, Aptdo. 22085, 46071, Valencia, Spain}

\address{$^3$Departament d'Estructura i Constituents de la Mat\`eria,
Universitat de Barcelona,
Diagonal 647, 08028 Barcelona, Spain}

\ead{tolos@kvi.nl}

\begin{abstract}
The properties of the $\bar K^*$ meson in dense matter are studied using a unitary approach in coupled channels within the framework of the local hidden gauge formalism. We obtain the $\bar K^*$ spectral function in the nuclear medium  and we found that the $\bar K^*$ develops an in-medium width up to five times bigger than in free space. We also estimate the transparency ratio of the $\gamma A \to K^+ K^{*-} A^\prime$ reaction, which we propose as a feasible experimental scenario  to detect in-medium modifications of the $\bar K^*$ meson.
\end{abstract}

\section{Introduction}
\vspace{0.5cm}
 The interaction of vector mesons with
nuclear matter has been a matter of investigation for years in connection to
fundamental aspects of QCD \cite{rapp,hayano,mosel}.
Many efforts have been invested in understanding the changes of $\rho$, $\omega$ and $\phi$ mesons in dense matter. The latest experiments conclude that there is no mass shift for the $\phi$ and $\rho$ in \cite{na60,wood} and for the case of the $\omega$ the conclusion in \cite{nanova} is that there is no evidence of a mass shift. 

However, no discussion has been made about the properties of the strange
vector mesons ($K^*$ and $\bar K^*$) in the medium. Only recently the $\bar K^*N$ interaction in free space has been addressed in  Ref.~\cite{GarciaRecio:2005hy} using SU(6) spin-flavour symmetry, and within the hidden local gauge formalism for the interaction of vector mesons with baryons of the octet \cite{Oset:2009vf} and the decuplet \cite{sarkar}.

The study of \cite{Oset:2009vf} allows us to obtain the $\bar{K}^*$ self-energy in the nuclear medium \cite{tolos10},
following a similar approach to the one employed in \cite{Ramos:1999ku,angels,Tolos:2006ny}. We shall find that there is a spectacular enhancement of the $\bar{K}^*$ width in the medium, up to five times the free value of 50 MeV. We also perform an estimate of the transparency ratio for 
$\bar{K}^*$ production in the $\gamma~A \to K^+ \bar{K}^{*-}A'$ reaction, which can be analyzed in present experimental facilities.

\section{$\bar K^*$ meson in dense matter}

\vspace{0.5cm}

The  $\bar K^*$  self-energy in symmetric nuclear matter \cite{tolos10} is obtained within the hidden gauge formalism. From this framework one can obtain the interaction of vector mesons among themselves \cite{hidden1,hidden2,hidden3,hidden4}. In particular, we extract a three-vector vertex term
\begin{equation}
{\cal L}^{(3V)}_{III}=ig\langle (V^\mu\partial_\nu V_\mu -\partial_\nu V_\mu
V^\mu) V^\nu\rangle
\label{l3V}\ ,
\end{equation}
with  $V_\mu$  the SU(3)
matrix of the vectors of the octet of the $\rho$ plus the SU(3) singlet.
One can also obtain the Lagrangian for the coupling of vector mesons to
the baryon octet given by
\cite{Klingl:1997kf,Palomar:2002hk}:
\begin{equation}
{\cal L}_{BBV} = g\left( \langle \bar{B}\gamma_{\mu}[V^{\mu},B]\rangle +
\langle \bar{B}\gamma_{\mu}B \rangle \langle V^{\mu}\rangle \right) ,
\label{lagr82}
\end{equation}
where $B$ is the SU(3) matrix of the baryon octet.
With these ingredients we can construct the Feynman diagrams that lead to  the vector-baryon ($VB$) transitions $VB
\to V^\prime B^\prime$. As discussed in Ref.~\cite{Oset:2009vf}, one can proceed
by neglecting the three momentum of the external vectors versus the vector
mass, in a similar way as done for chiral Lagrangians in the low energy
approximation, and one obtains the transition potential: 
\begin{equation}
V_{i j}= - C_{i j} \, \frac{1}{4 f^2} \, \left( k^0 + k^\prime{}^0\right)
~\vec{\epsilon}\,\vec{\epsilon }\,^\prime , \label{kernel} 
\end{equation} 
where $f$ is the pion decay constant, 
$k^0, k^\prime{}^0$ are the energies of the incoming and outgoing vector
mesons, respectively, $\vec{\epsilon}\,\vec{\epsilon }\,^\prime$ is the product of their
polarization vectors, and $C_{ij}$ are the channel coupling  coefficients
\cite{Oset:2009vf}.

The meson-baryon scattering amplitude is obtained from the coupled-channel
on-shell Bethe-Salpeter equation \cite{angels,ollerulf}
\begin{equation}
 T = [1 - V \, G]^{-1}\, V \ ,
\end{equation}
with $G$ being the loop
function of a vector meson and a baryon, which is conveniently regularized
 taking a natural value of $-2$ for the subtraction constants
at a regularization scale $\mu=630$ MeV \cite{Oset:2009vf,ollerulf}. Note that the relatively large decay width of the $\rho$ and  $\bar K^*$ vector mesons (into $\pi\pi$ or $\bar K\pi$ pairs, respectively)
is incorporated in the loop functions via the 
convolution of the $G$ function \cite{nagahiro}.

Since we are interested in studying the interaction of $\bar K^*$ mesons in nuclear matter, we concentrate in the strangeness $S=-1$ vector meson-baryon 
sector with isospin $I=0$ ($\bar K^*N$, $\omega \Lambda$, $\rho \Sigma$,
$\phi \Lambda$ and $K^* \Xi$) and $I=1$
($\bar K^*N$, $\rho \Lambda$, $\rho \Sigma$, $\omega \Sigma$, $K^* \Xi$ and
$\phi \Sigma$).  The factorization  of  the factor
$\vec{\epsilon}\,\vec{\epsilon }\,^\prime$ for the
external vector mesons also in the $T$ matrix \cite{Oset:2009vf} provides degenerate pairs of dynamically generated
resonances which have  $J^P=1/2^-,3/2^-$. 
For the $I=0$ and $I=1$ sectors two resonances are generated, 
 $\Lambda(1783)$ and  $\Sigma(1830)$, which can be identified with the experimentally observed states $J^P=1/2^-$ $\Lambda(1800)$ and $\Sigma(1750)$, respectively \cite{Oset:2009vf}.

Medium modifications on the $\bar K^* N$ scattering amplitude, which are incorporated in the $\bar K^*N$ loop function, come from two sources:  a) the contribution associated to the decay mode $\bar K \pi$ modified by the nuclear medium effects on the $\bar K$ and $\pi$ mesons, and b) the contribution associated to the interaction of the $\bar K^*$ with the nucleons in the medium, which accounts for the direct quasi-elastic process $\bar K^* N \to \bar K^* N$ as well as other absorption channels $\bar K^* N\to \rho Y, \omega Y, \phi Y, \dots$ with $Y=\Lambda,\Sigma$. 

\subsection{${\boldmath \bar{K}^*}$ self-energy from decay into ${\boldmath \bar{K}\pi}$}

\vspace{0.3cm}

The imaginary part of the free $\bar K^*$ self-energy at rest due to the decay of the $\bar{K}^*$ meson into $\bar{K}\pi$ pairs, ${\rm Im} \Pi^0_{\bar K^*}$, determines a value of the $K^{*-}$ width  of $\Gamma_{K^{*-}}=-\mathrm{Im}\Pi_{\bar{K}^*}^{0}/m_{\bar K^*}=42$ MeV \cite{tolos10}, which is quite close to the experimental value $\Gamma^{\rm exp}_{K^{*-}}=50.8\pm 0.9$ MeV.

\begin{figure}[t]
\begin{center}
\includegraphics[width=0.4\textwidth,height=4cm]{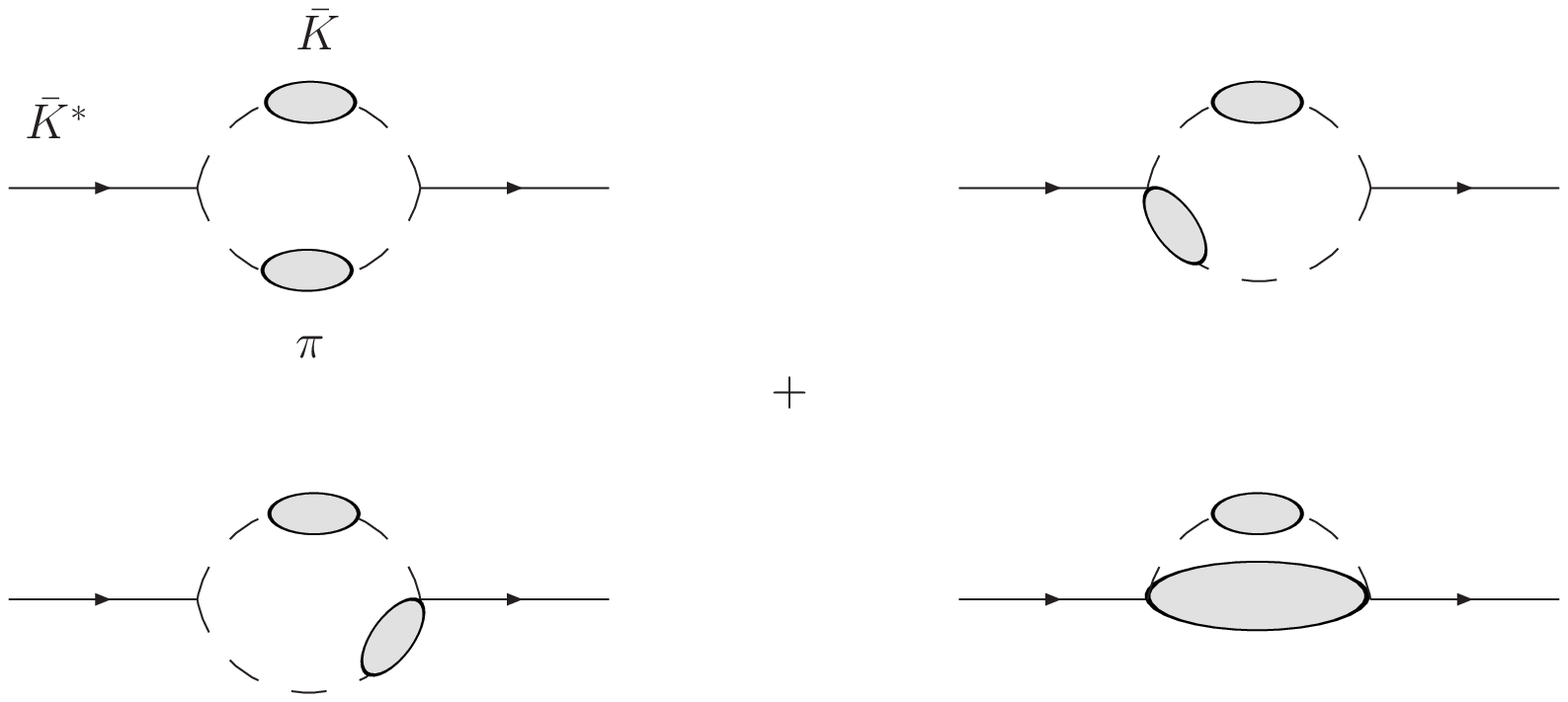}
\hfill
\includegraphics[width=0.5\textwidth,height=6cm]{Fig3_colouronline.eps}
\caption{Left: Self-energy diagrams contributing to the decay of the $\bar{K}^*$ meson in the medium. Right: Imaginary part of the $\bar K^*$ self-energy at $\vec{q}=0 \, {\rm MeV/c}$,
 coming from the ${\bar K}\pi$ decay mode in
dense matter at  $\rho_0$}
\label{fig:1}
\end{center}
\end{figure}

In the nuclear medium we shall calculate the $\bar{K}^*$ self-energy  coming from its
decay into ${\bar K}\pi$, $\Pi_{\bar{K}^*}^{\rho,{\rm (a)}}$, including both the self-energy of the antikaon \cite{Tolos:2006ny} and the pion \cite{Oset:1989ey,Ramos:1994xy} (see first diagram in the left hand side of Fig.~\ref{fig:1}). Moreover, one still has to implement vertex corrections, which are required by the gauge invariance of the model, and are associated to the last three diagrams in the l.h.s. of Fig. \ref{fig:1}.

In the right panel of Fig.~\ref{fig:1} we show the imaginary part of
the $\bar{K}^*$ self-energy for $\vec{q}=0$ coming from $\bar{K}\pi$ decay, in
free space (dotted line),  adding the $\pi$ self-energy (dot-dashed line)
and including both $\pi$ and
$\bar{K}$  self-energy contributions (dashed line)  at
normal nuclear matter saturation density ($\rho_0=0.17$ \ fm$^{-3}$). We can see 
that the main contribution to the $\bar K^*$ self-energy comes from the pion self-energy in dense matter. This is  due to  the fact that the 
$\bar K^*  \to \bar K \pi$ decay process leaves the pion with energy right in
the region of $\Delta N^{-1}$ excitations, where there is considerable pionic
strength. Vertex corrections (solid line) reduce the effect of the  pion dressing on the $\bar K^*$ self-energy, giving a $\bar K^*$ width of
$\Gamma_{\bar{K}^*}(\rho=\rho_0)=105$ MeV, which is about twice the value of the width in vacuum.

\subsection{$\bar K^*$ self-energy from the s-wave $\bar K^* N$ interaction}
\vspace{0.3cm}

We also consider the contributions to the $\bar K^*$ self-energy
coming from its interaction with the nucleons in the Fermi sea. These are affected by, on one hand, the Pauli principle acting on the nucleons, and on the other hand, the change of the properties of mesons and baryons in the coupled channel states
due to the interaction with nucleons of the Fermi sea. In particular, we
consider the self-consistently calculated $\bar K^*$ self-energy in the $\bar
K^*N$ intermediate states.

We solve the on-shell Bethe-Salpeter equation in nuclear matter for
the in-medium amplitudes in the isospin basis, $T^{\rho,I} = [1 - V^I \, G^{\rho}]^{-1}\, V^I$. The in-medium $\bar K^*$ self-energy is  then obtained by
integrating ${T^\rho}_{\bar K^*N}$ over the nucleon Fermi sea,
\begin{eqnarray}
\Pi_{\bar{K}^*}^{\rho,{\rm (b)}}(q^0,\vec{q}\,)&=&\int \frac{d^3p}{(2\pi)^3} \, n(\vec{p}\,)\,
\left [~{T^\rho}^{(I=0)}_{\bar K^*N}(P^0,\vec{P})+3 {T^\rho}^{(I=1)}_{\bar K^*N}(P^0,\vec{P})\right ] \ ,
 \label{eq:pid}
\end{eqnarray}
where $P^0=q^0+E_N(\vec{p}\,)$ and $\vec{P}=\vec{q}+\vec{p}$ are the
total energy and momentum of the $\bar K^*N$ pair in the nuclear
matter rest frame, and the values $(q^0,\vec{q}\,)$ stand for the
energy and momentum of the $\bar K^*$ meson also in this frame. The
self-energy $\Pi_{\bar{K}^*}^{\rho,{\rm (b)}}$ has to be determined self-consistently
since it is obtained from the in-medium amplitude
${T}^\rho_{\bar K^*N}$ which contains the $\bar K^*N$ loop function
${G}^\rho_{\bar K^*N}$, and this last quantity itself is a function of the complete self-energy
$\Pi_{\bar K^*}^{\rho}=\Pi_{\bar{K}^*}^{\rho,{\rm (a)}}
+\Pi_{\bar{K}^*}^{\rho,{\rm (b)}}$.

\subsection{Results for the $\bar K^*$ properties in dense matter}
\vspace{0.3cm}

\begin{figure}[t]
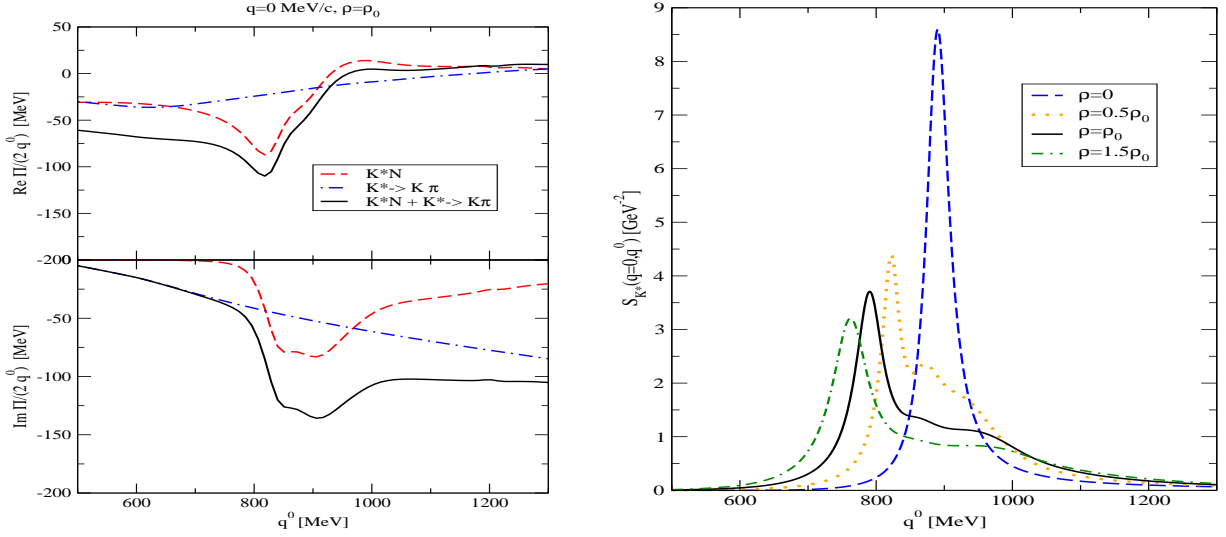

\begin{center}
\includegraphics[width=0.45\textwidth,height=7cm]{Fig5_colouronline.eps}
\hfill
\includegraphics[width=0.5\textwidth,height=7cm]{Fig7_colouronline.eps}
\caption{ Left: $\bar K^*$ self-energy for
 $\vec{q}=0 \, {\rm MeV/c}$ and $\rho_0$. Right: $\bar K^*$ spectral function for $\vec{q}=0 \, {\rm MeV/c}$  and different densities .}
\label{fig:auto-spec}
\end{center}
\end{figure}

We show next in the left hand side of Fig.~\ref{fig:auto-spec} the complete $\bar K^*$ self-energy as a function of the $\bar K^*$ energy $q^0$ for zero momentum at normal nuclear matter density. We display the contribution to the self-energy coming from the self-consistent calculation of the $\bar K^* N$ effective interaction (dashed lines) and the self-energy from the $\bar K^* \rightarrow \bar K \pi$ decay mechanism (dot-dashed lines), together with the combined result from both sources (solid lines).

For $\bar K^*$ energies around 800-900 MeV we observe an enhancement of the width together with some structures in the real part of the self-energy. These result from the coupling of the $\bar K^*$ to the dynamically generated $\Lambda(1783) N^{-1}$ and  $\Sigma(1830) N^{-1}$ excitations, which dominate the behavior of the $\bar K^*$ self-energy in this energy region. However, at lower energies where the $\bar K^* N\to V B$ channels 
are closed, or at large energies beyond the resonance-hole excitations,
the width of the $\bar K^*$ is governed by the $\bar K \pi$ decay mechanism in dense matter. 
At the
$\bar K^*$ mass, the $\bar K^*$ feels a moderately attractive optical potential and
acquires a width of $260$ MeV, which is about five times its width in vacuum.

The $\bar K^*$ meson spectral function, which results from the imaginary part of the in-medium $\bar K^*$ propagator, and it is given by
\begin{equation}
S_{\bar K^*}(q^0,\vec{q}\,)=-\frac{1}{\pi} \, {\rm Im} \, \left[ \frac{1}{(q^0)^2-(\vec{q}\,)^2-m_{\bar K^*}^2-\Pi_{\bar K^*}^{\rho}(q^0,\vec{q}\,)} \right] \ ,
\end{equation}
is displayed in the right panel of Fig.~\ref{fig:auto-spec} as a function of the meson energy $q^0$, for zero momentum and different densities up to 1.5 $\rho_0$. The dashed line refers to the calculation in free space, where only the $\bar K \pi$ decay channel contributes, while the other three lines correspond to fully self-consistent calculations, which also incorporate the process $\bar K^* \rightarrow \bar K \pi$ in the medium. 

 The structures above the quasiparticle peak correspond to $\Lambda(1783) N^{-1}$ and  $\Sigma(1830) N^{-1}$ excitations. Density effects result in a dilution and merging of those resonant-hole states, together with a general broadening of the spectral function  due to the increase of collisional and absorption processes. 
Although the real part of the optical potential is moderate, -50 MeV at $\rho_0$, the interferences with the resonant-hole modes push the $\bar{K}^*$  quasiparticle peak to even lower energies. However, transitions to
pseudoscalar-meson states that are not included in the model, such as $\bar K^*N \to \bar K N$, would make the peak 
less prominent and difficult to disentangle from the other excitations.
In any case, what is clear from the present approach, is that the 
width of the $\bar K^*$ increases substantially in the medium, becoming, at normal nuclear
matter density, five times bigger than in free
space.

\begin{figure}[t]
\begin{center}
\includegraphics[width=0.6\textwidth]{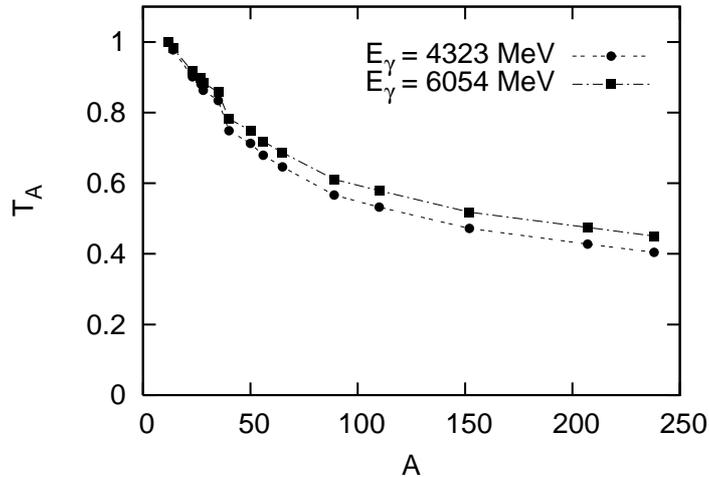}
\caption{Transparency ratio  for  $\gamma A \to K^+ K^{*-} A'$}
\label{fig:ratio}
\end{center}
\end{figure}

\section{Transparency ratio for  $\gamma A \to K^+ K^{*-} A'$}
\vspace{0.5cm}

 In order to test experimentally the $\bar K^*$ self-energy, we can study the nuclear transparency ratio by comparing the cross sections of the photoproduction reaction $\gamma A \to K^+ K^{*-} A'$ in different nuclei, and tracing the differences to the in medium $K^{*-}$ width.

The normalized nuclear transparency ratio is defined as
\begin{equation}
T_{A} = \frac{\tilde{T}_{A}}{\tilde{T}_{^{12}C}} \hspace{1cm} ,{\rm with} \ \tilde{T}_{A} = \frac{\sigma_{\gamma A \to K^+ ~K^{*-}~ A'}}{A \,\sigma_{\gamma N \to K^+ ~K^{*-}~N}} \ .
\end{equation}
The quantity $\tilde{T}_A$ is the ratio of the nuclear $K^{*-}$-photoproduction cross section
divided by $A$ times the same quantity on a free nucleon. It describes the loss of flux of $K^{*-}$ mesons in the nucleus and is related to the absorptive part of the $K^{*-}$-nucleus optical potential and, thus, to the $K^{*-}$ width in the nuclear medium.  We evaluate this ratio with respect to $^{12}$C, $T_A$, so that other nuclear effects not related to the absorption of the $K^{*-}$ cancel.

The results for different nuclei can be seen in  Fig. \ref{fig:ratio}, where the transparency ratio has been plotted for two different energies in the center of mass reference system, $\sqrt{s}=3$ GeV and $3.5$ GeV, which are equivalent to energies of the photon in the lab frame of $4.3$ GeV and $6$ GeV respectively. We observe a very strong attenuation of the $\bar{K}^*$ survival probability coming from the decay or absorption channels $\bar{K}^*\to \bar{K}\pi$ and $\bar{K}^*N\to \bar K^* N, \rho Y, \omega Y, \phi Y, \dots$, with increasing nuclear-mass number $A$. This is due to the larger path that the $\bar{K}^*$ has to follow before it leaves the nucleus, having then more chances to decay or being absorbed.

\section{Conclusions}
\vspace{0.5cm}

We have studied the properties of $\bar K^*$ mesons in symmetric
nuclear matter within a self-consistent coupled-channel
unitary approach using hidden-gauge local symmetry.
The corresponding in-medium solution incorporates Pauli blocking
effects and the $\bar K^*$ meson self-energy in a
self-consistent manner, the latter one including the $\bar K^* \to \bar K \pi$ decay in dense matter. In particular, we have obtained the self-energy and, hence, the spectral
function of the $\bar K^*$ meson.

We have found a remarkable change in
the $\bar K^*$ width in nuclear matter as compared to free space. At normal nuclear matter density the $\bar{K}^*$ width is found to be about $260$ MeV, five times larger than its free width. We have also made an estimate of the transparency ratio for different nuclei in the $\gamma A\to
K^+\bar{K}^* A'$ reaction and found a substantial reduction from unity of that
magnitude. The analysis of the transparency ratio is a very efficient experimental tool to study changes in the width of hadrons in dense matter. 

\section*{Acknowledgments}
\vspace{0.5cm}

L.T. acknowledges support from the RFF program of the University of Groningen. This work is partly supported by the EU contract No. MRTN-CT-2006-035482 (FLAVIAnet), DGICYT contract  FIS2006-03438 and
the Generalitat Valenciana in the program Prometeo.  We acknowledge the support of the European Community-Research Infrastructure Integrating Activity ``Study of Strongly Interacting Matter'' (HadronPhysics2, Grant Agreement n. 227431) under the 7th Framework Programme of EU.

\section*{References}
\vspace{0.5cm}

\end{document}